\documentclass{article}

\usepackage[english]{babel}

\usepackage[a4paper,top=2cm,bottom=2cm,left=3cm,right=3cm,marginparwidth=1.75cm]{geometry}

\usepackage{amsmath}
\usepackage{graphicx}
\usepackage[colorlinks=true, allcolors=blue]{hyperref}
\usepackage{comment}

\usepackage[backend=biber,style=nature, sorting=none,sortcites=true]{biblatex}
\usepackage{csquotes}
\usepackage{xcolor}
\usepackage{siunitx}

\usepackage{setspace}
\usepackage[normalem]{ulem}

\usepackage{authblk}

\usepackage{authblk}
\title{Direct observation of electron shedding from a laser-plasma accelerator}

\author[1]{Sheroy Tata}
\author[1]{Salome Benracassa}
\author[1]{Heychal Davidovich}
\author[1]{Aaron Liberman}
\author[1]{Anton Golovanov}
\author[1]{Eitan Levine}
\author[2]{Yang Wan}
\author[1]{Eyal Kroupp}
\author[1]{Victor Malka}

\affil[1]{Department of Physics of Complex Systems, Weizmann Institute of Science, Rehovot 7610001, Israel}
\affil[2]{Laboratory of Zhongyuan Light, School of Physics, Zhengzhou University, Zhengzhou, 450001, China}

\newif\ifParaSummary \ParaSummaryfalse 
\newif\ifSecMethods \SecMethodstrue 
\newif\ifSecFigures \SecFigurestrue 
\newif\ifSecRandom \SecRandomfalse 

\begin{document}

\maketitle

\textbf{
Laser–plasma accelerators have demonstrated the ability to produce ultrashort relativistic electron bunches with peak currents suitable for compact light sources, ultrafast diffraction, and strong-field studies. However, their performance critically depends on preserving the longitudinal phase-space density of the beam as it exits the plasma accelerator.
Here, we report the first direct observation of a previously unresolved process in which a highly charged electron bunch undergoes significant longitudinal expansion and progressively loses electrons during extraction from a laser-driven wakefield accelerator, a phenomenon we refer to as electron shedding. Using femtosecond relativistic electron microscopy, we tracked the evolution of the beam far beyond the accelerator exit and observed the bunch stretching over many plasma wavelengths before shedding electrons during propagation.
Particle-in-cell simulations reproduce the observed behavior and reveal that it originates from a combination of effects when a high-charge-density beam exits the accelerator. 
These dynamics redistribute energy from the beam head into a low-energy tail, thereby reducing the useful peak charge density and ultimately decreasing the efficiency of the laser–plasma accelerator. Our results provide new insight into beam extraction and phase-space evolution in laser plasma accelerators and highlight the importance of controlling these collective effects for future applications.
}


Ultrashort relativistic electron bunches are crucial for developing compact advanced light sources~\cite{schlenvoigt2008compact, wang2021free, labat2023seeded}, imaging ultrafast structural dynamics~\cite{Sciaini2011, wood2018ultrafast}, and testing fundamental physics~\cite{Marklund2009, Piazza2012}. In these applications, the spectral and temporal characteristics of the electron bunch directly influence the properties of the emitted radiation. To achieve femtosecond-duration light sources, the longitudinal phase space of the electron bunch must be confined to a few micrometers. In addition, the slice emittance and charge density are key parameters that must be optimized to advance these developments.
Laser-driven wakefield accelerators (LWFAs) are attractive candidates for achieving these objectives. They provide intense electron beams with kiloampere-level currents and source sizes limited to a few micrometers in all spatial dimensions~\cite{Corde2011, buck2011real, Schnell2012, Plateau2012, wan2023femtosecond}. Moreover, LWFAs can accelerate several hundred picocoulombs of charge using 100-TW-class lasers~\cite{irman2018improved, gotzfried2020physics}. The combination of short bunch duration and high charge offers a pathway to achieving the high charge density required for future light sources~\cite{wang2021free, labat2023seeded}. However, high charge density also introduces unique challenges in controlling the spatial and spectral distribution of the beam. It is therefore crucial to understand the mechanisms that affect the charge-density distribution as a high-charge beam exits an LWFA, in order to prevent degradation of performance.

In an LWFA, the bunch duration is typically limited by the length of the accelerating phase, which is on the order of the plasma wavelength~\cite{mangles2006laser, pukhov2002laser, guillaume2015physics}. 
Depending on the plasma density, this corresponds to several tens of micrometers, resulting in ultrashort bunches with durations ranging from a few to several tens of femtoseconds~\cite{lundh2011, lundh2013}. After acceleration, however, it is often assumed that the bunch length remains unchanged. This assumption is only valid when the target parameters are carefully tailored to prevent degradation of the temporal profile at the accelerator exit.
At the exit of an LWFA, the accelerated beam must traverse a plasma density downramp before transport into vacuum. 
In this transition region, the coupling between the plasma and vacuum strongly influences the transverse beam quality, as the downramp density profile can affect the beam divergence, emittance, and energy spread~\cite{sears2010emittance,floettmann2014adiabatic}. Tailored plasma profiles and plasma lenses have been shown to improve beam divergence and transport~\cite{van2015active, thaury2015demonstration, chang2023reduction}. However, far less attention has been devoted to understanding how this same transition reshapes the longitudinal charge distribution of a high-charge electron bunch, despite the critical importance of this quantity for many applications.

Experimental measurements of bunch duration from an LWFA have been carried out using coherent transition radiation (CTR) in the optical~\cite{lundh2011, lundh2013, lumpkin2020coherent, laberge2024revealing} and terahertz regimes~\cite{ADebus2010}, with plasma densities around \SI{e18}{cm^{-3}} resulting in femtosecond duration bunches. However, CTR is intrinsically most sensitive to longitudinal features shorter than the wavelength of measurement. As the bunch length increases, the emission from various slices add incoherently and results in loss of sensitivity and imposing an upper limit on the measurable bunch duration~\cite{bajlekov2013longitudinal}. Consequently, CTR preferentially reveals femtosecond-scale structures while underestimating or missing a continuously evolving, hundreds of micrometres long distribution with a low-energy tail. Earlier reports of few-femtosecond bunch durations, therefore do not necessarily exclude substantial longitudinal expansion during beam extraction.
Coherent optical transition radiation interferometry (COTRI) extends these measurements by recovering spectral phase through interference between two transition-radiation sources~\cite{lumpkin2020coherent,laberge2024revealing} has revealed trains of femtosecond bunches separated by approximately a plasma wavelength~\cite{heigoldt2015temporal}. Conversely, terahertz electro-optic measurements have indicated substantially longer trailing charge distributions~\cite{ADebus2010} although their temporal resolution is limited by the finite electro-optic response of materials, limiting their ability to measure short bunches. These techniques provide important measurements of the bunch structure at selected locations, but do not directly resolve its evolution as the beam exits the plasma.

In this paper, we show that electron shedding initiates in the density downramp at the exit of the accelerator. As the plasma density decreases, the plasma wavelength increases and the wake phase shifts relative to the driver, progressively moving the electrons into a decelerating phase of the wake. During this process, a fraction of the electrons lose energy and slow down to sub-luminal velocities, causing the bunch to elongate. Additional injection in the downramp further contributes to the process of lengthening. As the beam subsequently propagates in vacuum, these low-energy electrons are continuously lost from the bunch.
In this work, the term electron shedding is used to describe the experimentally observed reduction of the useful relativistic charge density during extraction of the beam from the plasma, caused by the formation of a low-energy, highly divergent tail that progressively separates from the main bunch. 
Electron shedding reduces the energy of the beam available for applications by modifying the spatial and spectral properties in the downramp. It also increases the beam dark current, which might lead to the formation of an electron halo surrounding the main bunch, as observed experimentally~\cite{kuschel2018controlling}.

Recently, femtosecond relativistic electron microscopy (FREM) has been used for characterizing linear and non-linear wakefields through their associated electric and magnetic fields~\cite{zhang2017femtosecond, wan2022direct} or the electron beams leaving a wakefield accelerator~\cite{wan2023femtosecond}.
In FREM, a probe electron beam propagates perpendicular to the probed wakefield and gains modulations in its transverse momenta from the fields in the plasma. After an additional drift, these momentum modulations evolve into density modulations, which are then imaged by impinging the probe onto a scintillating screen.
This technique provides spatial resolution of a few microns and temporal resolution of a few femtoseconds, making it a powerful method for capturing the dynamics of the wakefield and the accelerated beam in a single shot.

In our experiment, FREM images recorded at various positions downstream of the LWFA show an unexpected longitudinal lengthening of a high-charge electron beam as it exits the laser-driven wakefield accelerator. Following the acceleration stage, the electron bunch is observed to continuously lengthen, eventually reaching more than 40 times the plasma wavelength of the acceleration stage. As this elongated beam propagates through near-vacuum conditions, electrons are progressively shed along the way. Far from the acceleration region, the beam eventually evolves back into a single electron bunch, followed by a low-energy electron tail.
To explain this unexpected beam lengthening, we combine experimental observations with particle-in-cell simulations, which identify the different physical processes contributing to the effect.
It is important to note that the elongated beam observed with FREM extends over several hundred micrometers. Detecting such structures using CTR diagnostics would therefore require measurements in the hundreds of micrometer spectral range, making experimental implementation particularly challenging.

The experiment was performed with the $2 \times 100$\,TW HIGGINS laser at Weizmann Institute of Science. Two laser-plasma accelerators were operated simultaneously, one generating the FREM probe beam, with an energy of approximately $200$\,MeV, and the other generating the high-charge beam from the LWFA used in the study. A schematic of the experimental setup is shown in Fig.~\ref{fig:setup}. 
The accelerator in the study was driven by a laser focused with an f/35 off-axis parabolic mirror to a peak intensity of approximately $7 \times 10^{18}$\,W\,cm$^{-2}$. The plasma was produced in a 1:5 mm diameter supersonic gas jet with a density of $3$--$5 \times 10^{18}$\,cm$^{-3}$ with a 97\% He / 3\% N$_2$ gas mixture for ionization injection. The accelerator produced beams with approximately $550$\,pC of charge above $100$\,MeV, measured on a charge-calibrated electron spectrometer, while its fields were simultaneously imaged with FREM.
In a single shot, FREM captured snapshots of the field structure with a field of view of nearly $900$ $\mu$m. Scanning the probe beam across the gas jet enabled us to observe the evolution of the high charge beam as it exits the laser-driven wakefield accelerator.

The experimental data shown in Fig.~\ref{fig:setup}(c) are the raw images of the probe charge density distribution acquired at various positions after the end of the gas jet. The bright regions indicate higher probe electron density, while the darker regions indicate lower density. Each frame is cropped to a length of nearly $450$ $\mu$m, with the driving laser traveling from left to right of the image. 
As the imaging is performed in a low-density region of the gas jet where the wakefield is weak, the probe directly images the beam's field onto the scintillating screen.
Each frame tracks the driver's position at various distances from the end of the gas jet. 
The lengthening structure is expected to be the electron beam exiting the laser driven wakefield accelerator~\cite{wan2023femtosecond} with evolution that has not been observed directly.

To understand the snapshot generated by FREM, consider a long relativistic beam propagating in vacuum in the $z$ direction with a charge density $\rho(x,y,z)$ with current  $I(z) = \int_{x,y}j_z(x,y,z)$. 
Under certain approximations, when the length of the charge distribution is much longer than the length scale where the transversely traveling probe gains most of its transverse momentum, we can then show that the transverse width of the distribution in the probe image imprints the linear charge density, $\lambda(z) = \int_{x,y} \rho(x,y,z)$, onto the screen. For a relativistic beam propagating at the speed of light $c$, $\lambda(z)$ can be approximated to $I(z) / c$.  These approximations are valid if the current density is slowly varying compared to the length of the beam, as the deflection of probe electrons from various slices of the beam due to the magnetic field nearly cancels out as they contribute to opposing forces on the particles in the probe (see Supplementary Information). 
This imprint on the probe beam is also nearly independent of the transverse structure of the probed electron beam, as features smaller than the length scale where the probe gains its transverse momentum are averaged out since both the probe and the field structure move at nearly the speed of light perpendicular to each other.

The evolution shown in Fig.~\ref{fig:setup} reveals a dark region in the probe intensity distribution that progressively elongates from the head of the electron beam as it exits the wakefield accelerator. Farther downstream, the beam develops a distinct head--tail structure, with the tail becoming increasingly extended and separated from the main bunch. At distances of up to $8$\,mm from the exit of the gas jet, the probe reveals signatures of a beam-driven wakefield in a plasma with a density more than three orders of magnitude below that of the plateau region ($\leq 1\times10^{16}\,\mathrm{cm}^{-3}$) (see supplementary information).
The observed beam lengthening occurs after the plasma density has fallen below $\sim1\times10^{17}\,\mathrm{cm}^{-3}$, where the plasma wavelength is approximately $100\,\mu\mathrm{m}$. Over the preceding $1\,\mathrm{mm}$, the density rises rapidly to $1.2$--$1.7\times10^{18}\,\mathrm{cm}^{-3}$, defining the density downramp of the gas jet. The density profile in this region was inferred from measurements of the local plasma wavelength and is shown by the red points in Fig.~\ref{fig:setup}.

To understand the physics behind the observed lengthening of the electron beam, we analyze various possible physical contributions. For a relativistic electron bunch, the longitudinal space charge field alone is insufficient to cause the bunch to lengthen in the propagation distances seen in the experiment. 
Since most electrons are accelerated to relativistic energies, a process that leads to the formation of low-energy electrons is the evolution of the beam in the downramp of the gas jet, where the beam drives the wake, shifting the wake phase backward.
This evolving wake can decelerate part of the previously accelerated bunch while simultaneously trapping and accelerating additional electrons injected at the downramp.
In this process, an electron bunch confined to approximately $21\,\mu\mathrm{m}$ during the acceleration phase lengthens to several hundred micrometers, forming a tail that leads to the onset of electron shedding.

Particle-in-cell (PIC) simulations were performed to identify the mechanisms responsible for the observed beam evolution using the plasma density profile extracted from the experimental data, shown by the red dots in Fig.~\ref{fig:setup}. The simulation results are presented in Fig.~\ref{fig:SubLumWake}, where the color map represents the electron charge density at successive positions through the downramp and the solid line shows the on-axis electric field, $E_z$; negative values correspond to an accelerating field for electrons.
Figure~\ref{fig:SubLumWake}(a) shows the laser-driven wake near the end of the density plateau, with an electron bunch carrying approximately $680\,\mathrm{pC}$ above $100\,\mathrm{MeV}$ strongly loading the wake. As the laser propagates into the density downramp, the reduced plasma density weakens its relativistic self-guiding~\cite{sprangle1992propagation, wagner1997electron}, causing the laser to defocus and reducing its ability to drive a wake. At the same time, the increasing plasma wavelength and the strong beam loading allow the high-charge electron bunch to contribute increasingly to the formation of the wake, resulting in a transition into a combined laser and beam-driven wake, as shown in Fig.~\ref{fig:SubLumWake}(b).
During this transition, the wake phase shifts backward relative to the laser-driven case, moving part of the previously accelerated bunch into a decelerating field. Simulations with reduced beam charge confirm that the high-charge bunch is essential to enhance this backward phase shift, thereby increasing the energy loss of the beam head and redistributing its energy into the formation of a tail (see supplementary information).
The backward evolution of the wake also extends the trapping region through the downramp, resulting in the injection of several nanocoulombs of charge. These electrons are accelerated to approximately $50\,\mathrm{MeV}$ while exhibiting large oscillations around the axis. 
Finally, as the density decreases, the head of the beam-driven wake starts to decelerate these injected electrons, as seen in Fig.~\ref{fig:SubLumWake}(c). The resulting energy loss and large transverse oscillations about the axis increase electron divergence of the tail, ultimately leading to electron shedding.

Figure~\ref{fig:LengtheningMeanEnergy} shows the evolution of the longitudinal phase space through and beyond the density downramp, revealing the onset of electron shedding. We define the head as the electrons injected before the transition to a beam-driven wake, occupying the first $21\,\mu\mathrm{m}$ of the beam, and the tail as the remaining injected electrons behind the head. The longitudinal phase space of the strongly loaded wake at the end of the plateau is shown in Fig.~\ref{fig:LengtheningMeanEnergy}(a, b). As the laser begins to defocus and the wake transitions toward a beam-driven wake and additional electrons are injected into the tail, progressively increasing the longitudinal extent of the beam as seen in Fig.~\ref{fig:LengtheningMeanEnergy}(c--f). This injection continues up to approximately $2.5\,\mathrm{mm}$ after the beginning of the downramp.
Once the wake becomes predominantly beam-driven, electrons in the head experience stronger deceleration while electrons farther back in the tail continue to be accelerated. This energy redistribution progressively flattens the longitudinal phase space, with electrons of the tail near the head decreasing from approximately $50\,\mathrm{MeV}$ to $15\,\mathrm{MeV}$ over the following $1\,\mathrm{mm}$ of propagation as seen in Fig.~\ref{fig:LengtheningMeanEnergy}(d--f). Farther downstream, the tail also enters a decelerating field and loses energy. Combined with its larger transverse divergence, this causes the tail to progressively separate from the head of the beam as seen in Fig.~\ref{fig:LengtheningMeanEnergy}(g, h), with electrons of up to approximately $30\,\mathrm{MeV}$ ultimately being shed during propagation into vacuum. 
Tens of micrometers flat current profile due to downramp injection has been reported in~\cite{bulanov1998particle, schmid2010density, buck2013shock, xu2017high, hue2023beam} that differs from the long $250\, \mu \mathrm{m}$ beam observed here, indicating electron injection continues even as the density falls by 2.5 orders of magnitude.

Fig.~\ref{fig:EnergyHeadTail} shows the energy and charge of the beam by separating the electrons into a head-tail structure with the beam defined by all the electrons with energy $\geq 5 \, \mathrm{MeV}$ to separate them from the background plasma. Several distinct regions can be identified along the propagation axis.
Between $-3$ and $0\,\mathrm{mm}$, the energy of the beam head increases in the laser-driven wakefield accelerator. During this stage, the energy contained in the tail also increases due to oscillating electrons trailing the plasma wake, however these electrons remain localized within the background plasma.
Over the next $2\,\mathrm{mm}$, as the transition from a laser-driven wakefield to a beam-driven regime occurs, a strong injection of electrons in the tail appears, as seen by the sudden increase of the charge of the tail. Simultaneously, as described above, the head loses energy while the tail gains energy, indicating a transfer of energy from the head into the tail. Although the head contains lower charge, higher-energy electrons, the tail consists of a much larger charge at lower energies. 
Over this interval, simulations show that the tail gains approximately $0.1\,\mathrm{J}$ at the expense of energy from the head, thereby reducing the energy retained in the useful high-energy component of the beam by nearly $20\,\%$. This energy redistribution, therefore, reduces the overall efficiency of the laser-driven wakefield accelerator.
Further downstream, between $2$ and $4\,\mathrm{mm}$, the tail itself begins to lose energy as it enters the decelerating phase of the wake. As these electrons decelerate and diverge, both the energy and charge retained in the tail decrease, eventually leading to electron shedding.

The FREM images reconstructed from the PIC simulations are shown in Fig.~\ref{fig:FullSimProbe} and reproduce the features observed experimentally as seen in the left column. From the simulations, the longitudinal current, $I(z)$, is extracted for electrons of energy above $2$, $5$, and $10\,\mathrm{MeV}$ shown by the blue, orange, and green curves, respectively. As the beam enters a plasma of sufficiently low density, it can be seen that for a current computed for electrons above $2\,\mathrm{MeV}$, the width of the dark region in the probe image corresponds to $I(z)$ from the simulation. The current above $ 5 \, \mathrm {MeV} $ or $10\,\mathrm{MeV}$ is insufficient for reconstructing the probe images.
According to the model, as $I(z)$ flattens out, the width of the dark region in the FREM images remains nearly constant as seen in Fig.~\ref{fig:FullSimProbe} at $z = 3\,\mathrm{mm}$. 
The high frequency modulations in the probe image are an imprint of a discontinuous current profile as seen in both experiment and simulations (see supplementary information).
Consequently, FREM thus enables online, single-shot estimation of the longitudinal charge distribution and the current profile of a long relativistic bunch.
As the beam propagates further, electrons are seen to shed from the beam.

In this work, by directly imaging the evolution in the longitudinal current of a high-charge electron beam exiting a laser wakefield accelerator, we have identified a previously unresolved process of electron shedding. As the beam propagates through the density downramp, the wake transitions from a laser driven to a beam driven wake, resulting in additional injection, acceleration, and deceleration of electrons, resulting in the formation of a long extended tail. The tail is subsequently lost as it diverges and separates from the head, progressively reducing the energy retained in the useful high energy component of the beam.
Our results show that electron shedding is governed by the coupling of a high charge beam as it exits the LWFA through the density downramp. The downramp therefore plays a critical role not only in preserving the transverse beam during extraction but also the longitudinal phase space. A gradual transition in the downramp promotes continuous injection and energy redistribution from the head into the tail, whereas a sharper transition suppresses these effects at the expense of increased beam divergence. Optimizing the plasma downramp for transport of a high charge density beam requires carefully balancing the preservation of the longitudinal phase-space against transverse beam divergence for their use in future applications.

\ifSecFigures
\begin{figure}
\centering
\includegraphics[width=\linewidth]{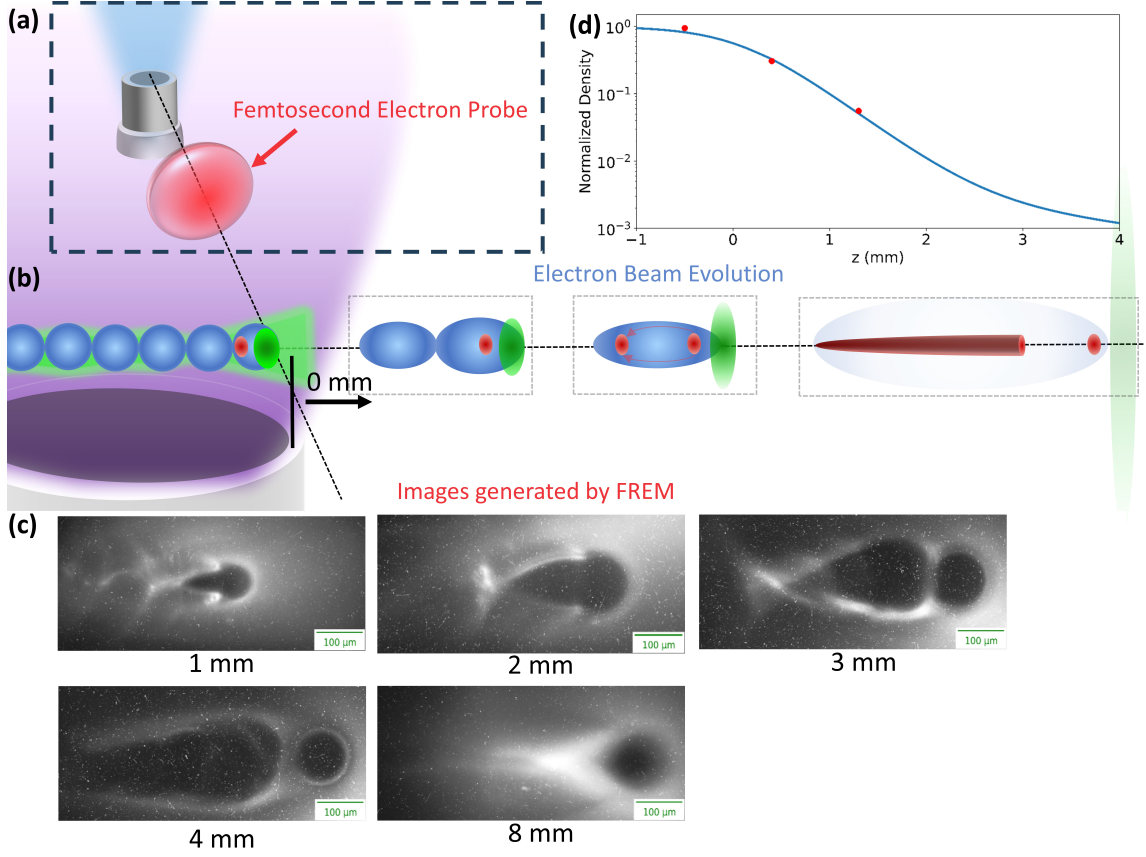}
\caption{\label{fig:setup}A Schematic of the setup, (a) a probe beam generated by an LWFA nearly perpendicular to the (b) LWFA being probed. The distances used in the data are from the downramp of the gas jet. (c) The evolution of the electron beam captured with FREM. Experimental data suggest lengthening and shedding of the beam as it exits the LWFA. The beam is seen to stretch to multiple plasma periods before a single beam is seen to propagate. Each of the frames is nearly $550~ \mu m$ long. (d) The density profile around the downramp of the gas jet, with the points in red are measurements of density from FREM, while the curve in blue is the density profile used for simulations.}
\end{figure}

\begin{figure}
\centering
\includegraphics[width=\linewidth]{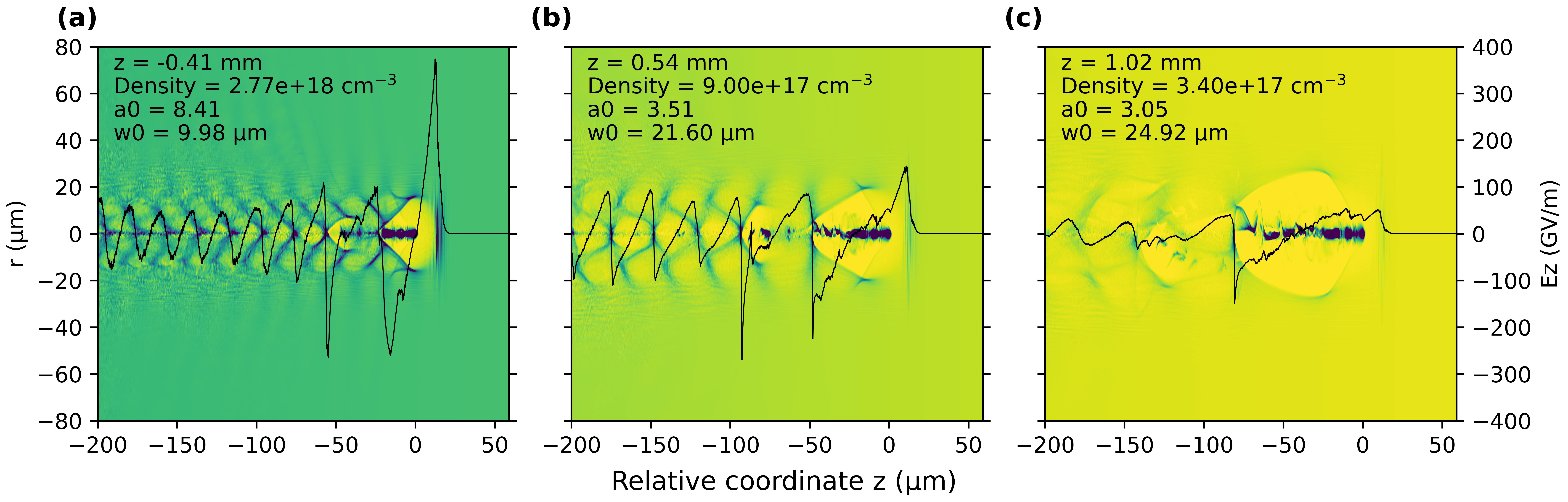}
\caption{\label{fig:SubLumWake} The panels show the evolution of the wakefield at the end of the LWFA. The solid line represents the $E_z$ field of the wake. (a) shows the wake as the plasma density falls and the beam head starts to enter the decelerating phase of the wake and starts to drive the wake. (b) represents a snapshot where the contribution due to the laser-driven wake falls, and the beam and the laser continue to drive a wake simultaneously. During this process, the back of the wake moves with a sub-luminal velocity, assisting injection at the downramp. (c) shows the laser-driven wake sufficiently weak and the beam driving the wake. Note that injection continues as the wake is driven by the initial LWFA-generated beam.}
\end{figure}

\begin{figure}
\centering
\includegraphics[width=\linewidth]{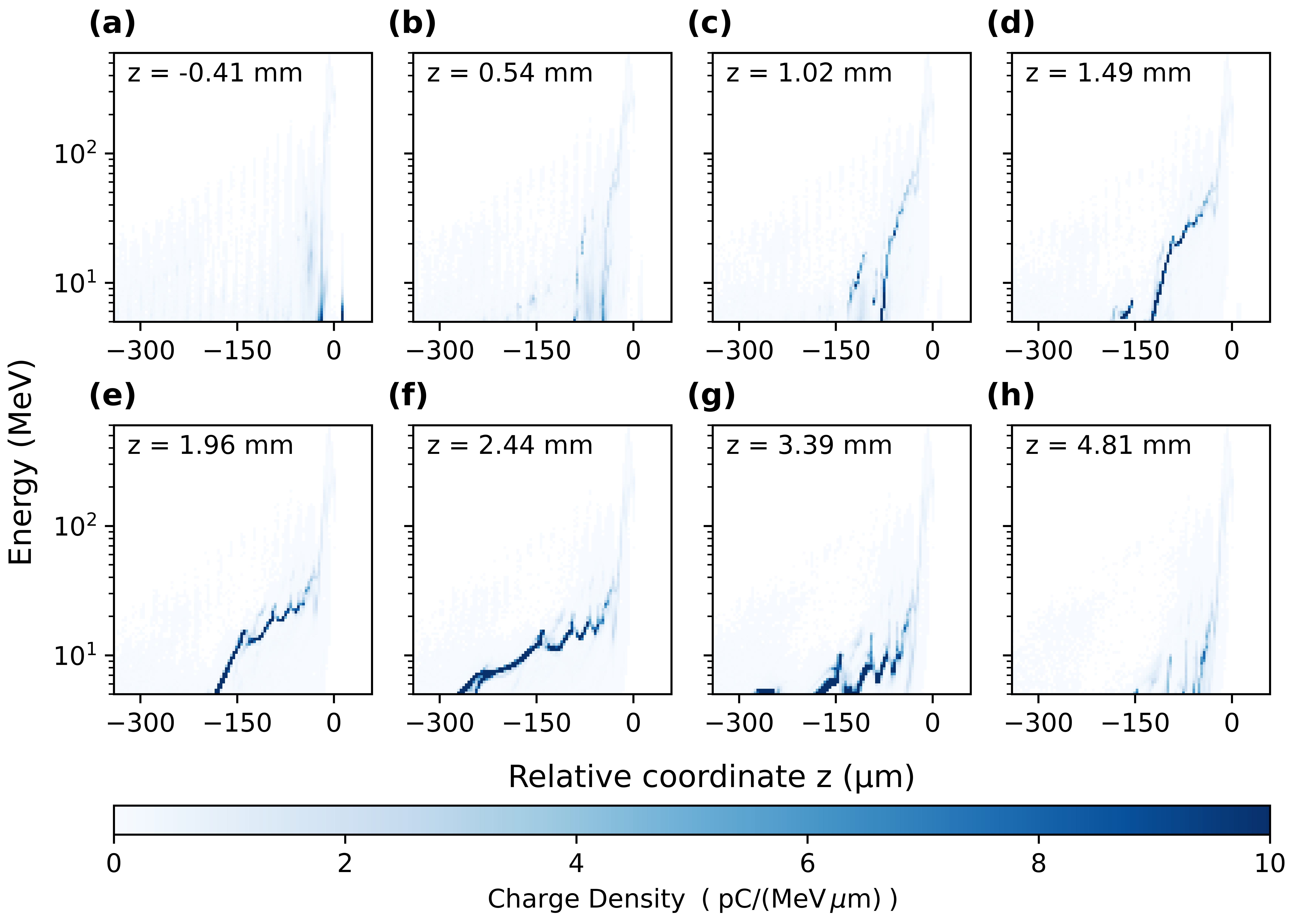}
\caption{\label{fig:LengtheningMeanEnergy} The evolution in the longitudinal phase space showing the lengthening and subsequent shedding of the electron beam. (a) represents a the phase space of a strongly loaded laser driven wake with electrons only in the head of the beam. (b -- d) As the plasma density decreases, the wake transitions into a beam driven regime resulting in strong injection and transfer of energy into driving the wake as seen by the energy loss near the head. (e, f) Injection continues across the downramp and energy is redistributed from near the head into the tail. (g, h) Farther downstream, all the electrons in the tail enter the decelerating field and increased divergence causes shedding of the beam.  }
\end{figure}

\begin{figure}
\centering
\includegraphics[width=\linewidth]{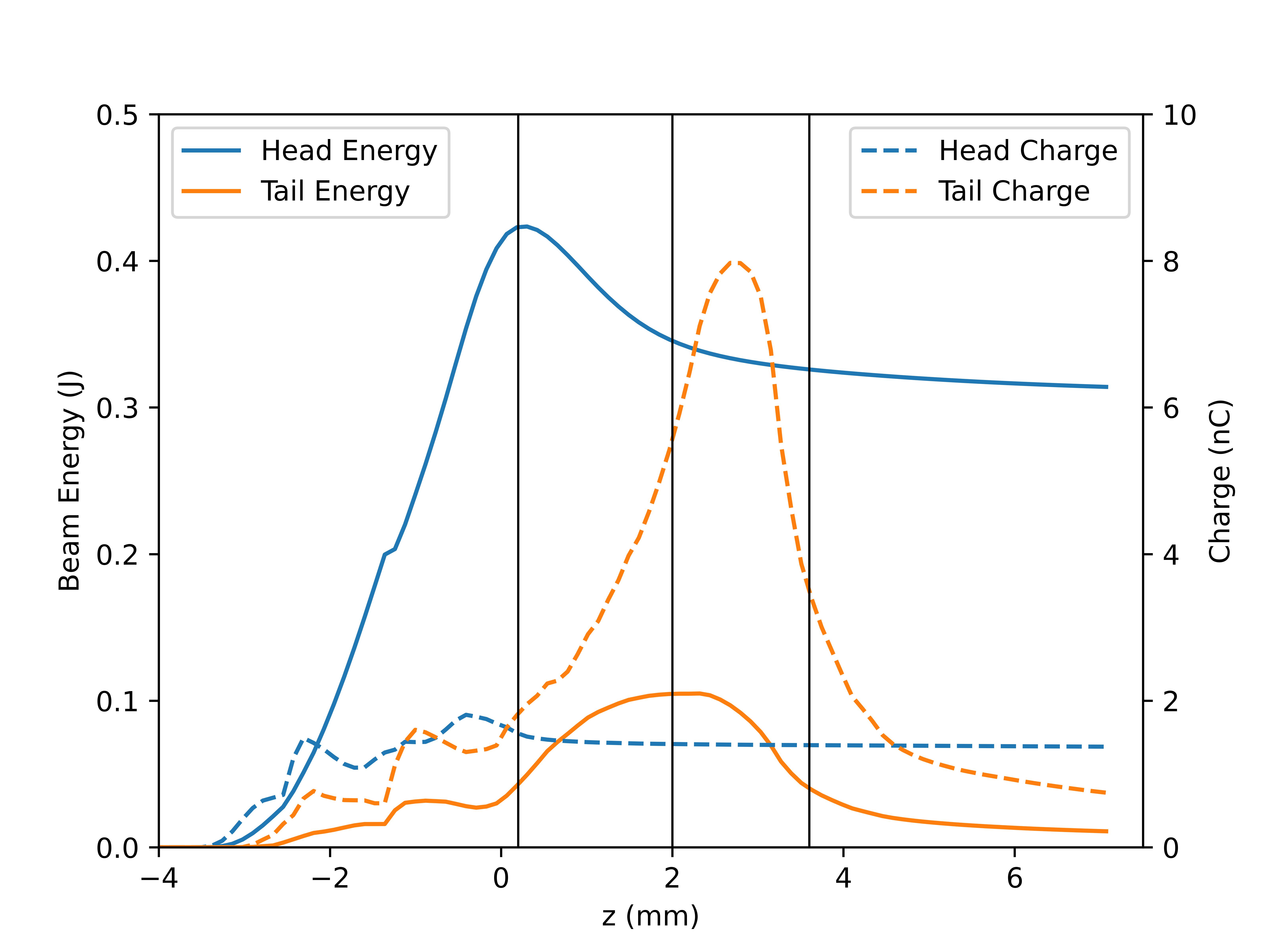}
\caption{\label{fig:EnergyHeadTail} Separating the beam into a head-tail structure and integrating the energy and charge of electrons with energy  above $5\,\mathrm{MeV} $. 
The head is defined as the first $21\,\mu\mathrm{m}$ of the beam, and the tail as the remaining part.
We present the energy evolution of the head and tail as the laser-driven wake propagates in plasma. Various stages are separated by the vertical lines, acceleration stage on the left and electrons in the laser-driven wakefield gain energy. After a certain point the charge in the head remains nearly constant. A small energy increase is observed in the tail due to background plasma electrons. In the subsequent stage, at the downramp, a strong injection of charge appears, with a significant part of the energy transferred from the head into the tail. Further downstream, the tail energy and charge is lost due to and deceleration and shedding of electrons from the beam. Finally on the right only the head of the beam survives with the tail carrying a fraction of the energy.}

\end{figure}

\begin{figure}
\centering
\includegraphics[width=\linewidth]{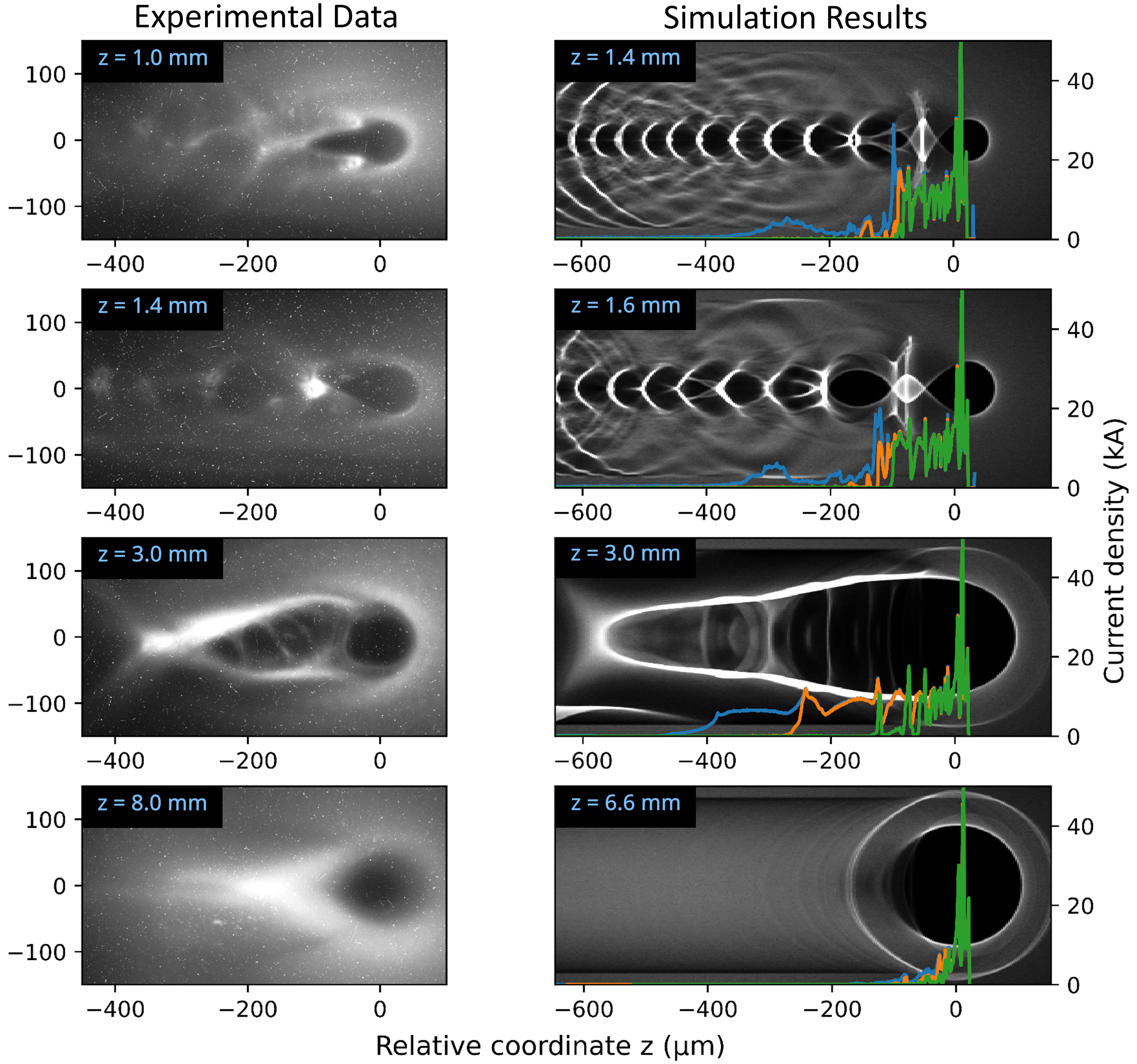}
\caption{\label{fig:FullSimProbe} Comparison between experimental and simulated evolution in FREM images of an high charge electron beam exiting a laser-driven wakefield accelerator. The left panel shows the raw data acquired by femtosecond relativistic electron microscopy of an electron beam leaving a laser driven wakefield accelerator. The panel on the right shows the images generated by simulations using the field and a quasi-static evolution, revealing processes that occur in an LWFA. 
The solid lines represent the current of the beam computed for energies above $2$\,,\,$5$\,,\,$10\,\mathrm{MeV}$ represented by the blue, orange and the green curves, respectively. It's important to note that even electrons near $2\,\mathrm{MeV}$ are necessary to recreate the longitudinal current of the beam as seen in the experiment, which are generated by various processes described in the article.}
\end{figure}
\fi

\ifSecMethods
\section*{Methods}

\subsection*{Generation of the probe beam and imaging the fields}
The electron probe beam was generated by an LWFA driven by a $1.7\,\mathrm{J}$, $30\,\mathrm{fs}$ laser pulse focused with a $1.5\,\mathrm{m}$ focal length off-axis parabola. Ionization injection in a $1:3\,\mathrm{mm}$ diameter supersonic gas jet containing a $97:3$\,\% mixture of He:N$_2$ was used to produce electron beams with energies of approximately $200\,\mathrm{MeV}$.
A motorized dielectric mirror positioned after the parabola was used to point the probe beam toward different positions of the LWFA under study. The probe beam energy was optimized before the experiment using an electron spectrometer with a movable magnet. The probe source was located approximately $11\,\mathrm{cm}$ from the probed gas jet, resulting in a probe diameter of approximately $1\,\mathrm{mm}$ at the region of interaction.
After interaction with the probed wakefield, the transverse probe-density modulation were imprinted on a $100\,\mu\mathrm{m}$ thick Ce:YAG screen positioned $16\,\mathrm{mm}$ after of the probed LWFA. The screen was imaged onto a CMOS camera using an imaging system with a field of view of approximately $900\,\mu\mathrm{m}$. The complete detection assembly could be translated along the propagation direction to probe different positions of the LWFA under study. To suppress laser from the probe imaging system, a $100\,\mu\mathrm{m}$ thick stainless steel foil was placed in front of the Ce:YAG screen, while an additional $50\,\mu\mathrm{m}$-thick black-anodized aluminium foil was used to reduce reflections back into the imaging system.

\subsection*{PIC simulations}
Particle in cell simulations were carried out using FBPIC. A size of 778 $\times$ 270 $\mu$m  was used for the z and r coordinates, respectively, with 16384 by 900 in each direction. There were two angular modes with 2, 2, 6 particles per cell in the $z$, $r$, and $\theta$ directions, with 16 points per wavelength. The normalized laser vector potential of $a_0 = 2$ with a $27~\mu m$ as the full width half maximum in intensity. A mixture of helium and nitrogen was used with helium being fully ionized and nitrogen pre-ionized up to level 5. The field outputs from the simulations were used to reconstruct the images generated by the electron probe. Since the simulations were large, the domain was decomposed into 4 equal sections, with 32 cells as the interpolation order used in the field solver. 

\subsection*{Reconstructing probe images from simulations}
Using the results from the PIC simulations, it is possible to reconstruct the images seen on the scintillating screen. Particle tracking of probe electrons was performed in a quasi-static evolving structure. The transverse momentum gained locally by the probe was used to compute the position after a drift. This particle density was
w projected onto a screen imprinting the field structure.

\fi

\section*{Author Contributions}
The experiment was carried out by ST, SB, HD, AL, EYL. The experiment was conceptualized by ST and VM. YW contributed to the code to reconstruct the probe images. ST, AG, VM wrote the paper. ST carried out simulations with discussions with AG. EK and VM provided support for the experiment. 

\section*{Funding}
The research was supported by the Schwartz/Reisman Center for Intense Laser Physics, the Benoziyo Endowment Fund for the Advancement of Science, the Israel Science Foundation (contracts 2412/22 and 1267/24), the Wolfson Foundation, the Schilling Foundation, and R. Lapon, Dita, and Yehuda Bronicki.

\printbibliography

\end{document}